\documentstyle[aps,epsf,multicol]{revtex}
\newcommand{\be}{\begin{equation}}
\newcommand{\ee}{\end{equation}}
\begin{document}
\preprint{SNUTP 97-86}
\draft
\title{Genetic Polymorphism in Evolving Population\\} 
\author {H.Y. Lee, D. Kim, and M.Y. Choi\\}
\address {Department of Physics and Center for Theoretical Physics\\ 
Seoul National University\\
 Seoul 151-742, Korea\\} 

\maketitle

\begin{abstract}
We present a model for evolving population which 
maintains genetic polymorphism. 
By introducing random mutation in the model population 
at a constant rate,
we observe that the population does not become extinct but survives, 
keeping diversity in the gene pool under abrupt environmental changes.
The model provides reasonable estimates for the proportions of 
polymorphic and heterozygous loci and for the mutation rate, as
observed in nature.
\end{abstract} 

          
\begin{multicols}{2}
\narrowtext 

 In the biological evolution process, it is known that the population
remains polymorphic, consisting of two or more genotypes: Genetic
variation thus persists \cite{Futuyma,Snyder}.
Balanced polymorphism stands for 
that the population consists of two or more genotypes with
the rate of the most frequent allele less than 95$\%$.
The proportion of polymorphic loci, measured by electrophoresis at allozyme
loci in animal and plants, takes the values in the range between 0.145 and
0.587 \cite{Selander,Hamrick}.
Provided that one allele replicates faster than others,
it increases in frequency at the expense of the others.
It may completely replace other alleles, and make the population
monomorphic, consisting of one genotype only.
Often, however, the allele does not completely replace others but remain at 
a stable intermediate frequency, leaving the population polymorphic,
although in nature the superficial similarity conceals 
the diversity of genotypes occurring among individuals within species.
This is of importance in the light of the fact that
the superiority of an allele holds only
in certain environments, or at certain gene frequencies, or in conjunction
with certain other alleles.
In biology, how to maintain polymorphism is 
described in a phenomenological manner by introducing 
such assumptions as the balance of selection and 
recurrent mutation,
selection balanced by gene flow, heterozygous advantage selection,
frequency-dependent selection, and variable selection 
in time or space \cite{Futuyma}. 

 Recently, a number of models for evolving population, including geographic
speciation and conditions for adaptation of population,
have been developed through the method of statistical mechanics 
\cite{Higgs,Mroz}.
In particular, Mr\'{o}z {\it et al.} have introduced 
an interesting population model in which
individuals are represented by their genotypes and phenotypes
while the environment characterizes the ideal phenotype\cite{Mroz}.
In this model, a random population must have
the initial adaptation greater than a certain critical value in order
to grow in an environment which does not change with time.  
In this paper, we attempt to 
describe genetic polymorphism, especially
under the environmental change,
by means of a population model which exhibits genetic polymorphism
within itself. It is shown that the population which 
displays genetically polymorphic
behavior is indeed advantageous for survival in the changing environment.
 
We consider a population consisting 
of $M$ individuals, each characterized by its sequence of genes 
with the size of the sequence given by $N$.
Each gene is assumed to take two possible forms, i.e., there are two alleles
denoted by $A$ and $a$.
Following Ref.~\cite {Mroz}, we consider diploidal organisms, 
in which the genotype at locus $\alpha~(=1,2,...,N)$ 
in the sequence can be $AA$ (represented by $G^{\alpha}=1$), 
$Aa=aA~(G^{\alpha}=0)$, or $aa~(G^{\alpha}=-1)$.
The genotype of the $i$th individual $(i=1,2,...,M)$ 
is then given by the sequence $\{G_i^{\alpha}\} 
\equiv \{G_i^1,G_i^2, ..., G_i^N\}$. 
We assume that the phenotype $F_i$ of the $i$th individual
simply follows its genotype with allele $A$ dominant:
If $G^{\alpha}_i=1$ or 0, then
$F^{\alpha}_i=1$; if $G^{\alpha}_i=-1$, then $F^{\alpha}_i=0$. 
The environment is characterized by a certain phenotype $\hat{F}$,
which is also represented by a sequence of $N$ numbers,
each of which being either 0 or 1.
We denote by $m$ the relative number of 1's in the ideal phenotype
$\hat{F}$.
The survival probability $p_i$ for the $i$th individual is 
assumed to be 
determined by the similarity of its phenotype to the ideal one $\hat{F}$:
\begin{equation}
p_i=N^{-1}\sum_{\alpha=1}^{N} \delta(F^{\alpha}_i,\hat{F}^{\alpha}),
\end{equation}
where $\delta(F^{\alpha}_i,\hat{F}^{\alpha})$ is the Kronecker delta symbol.
The average adaptation of the population to the environment 
is defined to be
\begin{equation}
\bar{A} \equiv M^{-1}\sum_{i=1}^{M}p_i.
\end{equation}
We begin with $M=M_0$ individuals with randomly distributed genotypes,
and generalize the adaptation model in Ref. \cite {Mroz} as follows:
The population is not allowed to grow indefinitely;
the maximum size of the population is set to be $L$. We first
choose randomly
a pair of individuals $i$ and $j$, and calculate survival probabilities
of these two individuals from Eq.~(1).
They may die at this stage with probability $1-p_i$ and $1-p_j$, respectively,
decreasing the value of $M$.
In case that both the two survive, 
we calculate the probability for the population size
\begin{equation}
P_p \equiv 1-M/L,
\end{equation}
according to which the individual $i$ and its mating partner $j$ become
parents.
Thus the probability $P_p$ provides the limit of the population size,
which is in contrast with the existing adaptation model \cite{Mroz}.
The parents produce $\gamma$ offsprings and then die.
The genotypes of the offsprings are determined from those of parents,
following the standard genetic rule.

Before considering the effects of environmental changes and mutations based 
on this model, we first examine the general behavior of the model.
With the initial population which has randomly distributed genotypes,
we examine the relation in the stationary state between $m$,
the density of 1's in $\hat{F}$, 
and $C \equiv M/L$, the concentration of the surviving
population.
As the size $N$ of the sequence increases, the concentration $C$
as a function of $m$ is found to approach a step-like structure,
which is similar to the one found in the existing adaptation model \cite{Mroz}.

It is further observed that $m_c$, which denotes
the the value of $m$
corresponding to the concentration of the
surviving population $C=0.5$,
depends not only on the size $N$
but also on the number of offsprings:
For given $N$, $m_{c}$ gets smaller/larger
as the number $\gamma$ of offsprings is increased/decreased.
For example, in case $N$=15,
we have $m_{c}=1$ for  $\gamma \leq 2$,
and $m_{c}=0$ for $\gamma \geq 6$.
As the size of the sequence grows, the minimum number of offsprings
required to keep the population from extinction, i.e.,
to have $m_{c}<1$, tends to increase.
Namely, for a given number of offsprings, the population eventually
becomes extinct as $N$ is increased;
this is found to be independent of
the initial concentration of the population.
This phenomenon is similar to the error threshold in
the quasispecies model\cite{Eigen}.
%
For given $N$ and $C$,
the minimum number of offsprings is also found to decrease
as the maximum size $L$ of the population grows.
This implies that the population with a larger value of $L$
can survive better than that with a smaller value, 
which reflects that large $M_0$ leads to less dispersion in 
the distribution of the average adaptation.

{\it Effects of environmental changes}: It is in general expected 
that the biological environment 
does not remain unchanged for a long time,
making it desirable to introduce change of the environment.
For this purpose, we choose randomly $n$ loci 
among the total $N$ loci of $\hat{F}$  
at every $10^6$ generations, and change their values.
Namely, if the selected ideal phenotype is 1,
it is changed to 0, and vice versa.
Figure 1(a) shows that the total population becomes extinct
after three times of the environmental change,
while the average adaptation vanishes as shown in Fig.~1(b). 
Such behavior is closely linked to genetic polymorphism.
To examine this, 
we define the proportion of polymorphic loci as follows:
\begin{equation}
P_{poly} \equiv \frac{1}{N}\sum_{\alpha=1}^{N}\chi_{\alpha},  
\end{equation}
where
$$
\chi_{\alpha} \equiv \left\{
\begin{array}{ll}
1 & \mbox{if    } (1/M)|\sum_{i=1}^{M} G_i^{\alpha}| \leq 0.9; \\
0 & \mbox{otherwise}.
\end{array}
\right. 
$$
Similarly,
the proportion of heterozygous loci, which
have genotype $Aa$, is defined by
\begin{equation}
P_{hetero} \equiv \frac{1}{M}\sum_{i=1}^{M}\sum_{\alpha=1}^{N} 
[1-{(G_{i}^{\alpha})}^2].
\end{equation}
The proportion of polymorphic loci in natural population is  
known to take the values between 0.145 and
0.587 while that of heterozygous loci lies in the range between 
0.037 and 0.170 \cite{Selander,Hamrick}.
Figure 1(c) exhibits that $P_{\it poly}$ does not keep its value but
decreases when the environmental change takes place. 
This implies that the population adapted to the environment initially 
becomes more 
monomorphic as the environment changes.
Figure 1(d) also shows that the proportion of heterozygous loci goes  
below the value in natural population as the environment changes.

This monomorphic tendency can be understood in terms of the  
Hardy-Weinberg theorem \cite{Futuyma}, which is applicable in the
absence of correlation between the loci in a single gene sequence.
In this case, each locus in the sequence may be considered independent. 
It is further assumed that the population is effectively infinite in size and 
that all individuals in the population mate at random.
Suppose that the gene frequency of allele $A$ is $p$ and that of 
allele $a$ is 
$q~ (\equiv 1 - p)$. Since mating is random within the population, 
the frequencies
of the genotypes $AA$, $Aa$, and $aa$ will be $p^2$, $2pq$, and $q^2$, 
respectively; this leads to
the frequency of $A$ among these progeny given by $p'=p^2+(2pq)/2=p$,
which is the same as the allele frequency in the previous generation.
On the other hand, when one takes into account the different fitnesses 
relative to the environment, the allele frequency in general changes with 
the generation.
Recall that genotypes $AA$ and $Aa$ have the phenotype $F=1$, with the  
fitness given by $m$,
the proportion of 1's in the ideal phenotype, whereas
the fitness of genotype $aa$ is given by $1-m$.
This leads, after one generation, to the
gene frequency for allele $a$:  
\begin{equation}
q'=\frac{mpq+(1-m)q^2}{m p^2+2 m p q +(1-m) q^2},
\end{equation}
and the change in the allele frequency per generation is given by
\begin{equation}
\Delta q \equiv q'-q 
        = - \frac{(2-m^{-1})pq^2}{1-(2-m^{-1})q^2},
\end{equation}
which is less than zero for $m > 1/2$. 
The stationary state condition $\Delta q =0$ then yields the stationary
frequency $q=0$; this implies that allele $a$ is replaced by $A$
which is more adaptive
to the environment. 

To prevent such monomorphic behavior, the system needs mutation.
Suppose that allele $A$ mutates to 
$a$ and vice versa, both with the mutation rate $u$.
The change in the allele frequency of $a$, due to mutation, per generation
is then given by
\begin{equation}
\Delta q_{\it mut}=u(1-q)-uq=u(1-2q),
\end{equation}
which leads to the total change in the allele frequency of $a$ per generation 
\begin{equation}
\Delta q_{tot} \equiv \Delta q+\Delta q_{mut}=
-\frac{(2-m^{-1})pq^2}{1-(2-m^{-1})q^2}+u(1-2q).
\end{equation}
Note that $\Delta q_{tot}$ can be either positive or negative and that
$\Delta q_{tot}=0$ in equilibrium gives
\begin{equation}
(2-m^{-1})[(1+2u)q_{eq}^3-(1+u)q_{eq}^2]-2uq_{eq}+u=0,
\end{equation}
the solution $q_{eq}$ of which is displayed in Fig.~2 
as a function of $u$, for various values of $m$.
It shows that the equilibrium value $q_{eq}$ increases from zero, regardless of $m$,
as the mutation rate $u$ grows.
In this way, polymorphism is easily achieved in the 
Hardy-Weinberg theorem if the effect of mutation is 
explicitly taken into account. 

{\it Roles of mutation}: We therefore consider mutation also in our model, 
which keeps
the population from going monomorphic, and hence prevents 
total extinction of the population. 
Mutation is known to occur at random in the sense
that the change of specific mutation is not affected by how useful that
mutation would be in the given environment \cite{Dobzhansky}.  
We thus add the mutation mechanism to the generation of population as follows:
Every $m_g$ generations, we select
one individual randomly, choose one allele in a randomly selected locus 
in the individual, and alter the allele. Namely, if the selected allele
is $A$, it is replaced by $a$, and vice versa.   
The mutation rate per generation per individual per locus is given by
$(m_g M N)^{-1}$. 

We have first examined the relation between $m$ and $C$ 
in the presence of mutation ($m_g \ne 0$),
and found that the 
random mutation process does not alter qualitatively the behavior of our model.
We have further investigated the effects of mutation on such quantities
as the concentration of the population,
the average adaptation, the proportion of polymorphic loci, 
and the proportion of heterozygous loci.
We have thus measured them at time $10^7$, 
after ten environmental changes,
each of which consists of changing $n{=}4$ loci among the total 
$N{=}15$ loci of the ideal phenotype. 
Figure 3 displays the behavior of those quantities 
with the mutation interval $m_g$.
Below the mutation interval $m_g=3$ (not seen in the scale of Fig.~3),
the population cannot survive the 
environmental change.
Thus too frequent mutations also
prevent the population from adapting to the environment.
As the mutation interval is increased, 
both the concentration of the population and the 
average adaptation becomes larger, approaching rapidly the value unity,
similarly to the case of the adaptation model
without mutation.
However, there exists a critical value of the mutation interval, 
beyond which the adaptation drops to zero
and the population becomes extinct completely. 
In Fig.\ 3(c) and (d), the proportions of polymorphic loci
and of heterozygous loci are shown to have the values around 0.4 and 0.1,
respectively, in the appropriate range of the mutation interval. 
These values of $P_{poly}$ and $P_{hetero}$, in the range
of the mutation interval for survival, are consistent with
the genetic variations observed in natural population. 
 
We have also considered dependence on the values of $n$,
the number of ideal phenotype loci changed when environment changes.
Figure 4 shows $\Delta m_g$, the range of the mutation interval 
for survival of population, as a function of $n$.
As the value of $n$ grows,
$\Delta m_g$ is shown to decrease exponentially.
When $n$ is larger than 6, which corresponds to the environmental change
greater than 40$\%$, the population cannot adapt to the 
changing environment and becomes extinct in spite of mutation. 
It is known from the study of individual proteins 
that mutation at individual loci in general 
arises at the rate of $10^{-6}$ to $10^{-5}$ per generation \cite{Mukai}.
The proper mutation rate $(m_g M N)^{-1}$, measured in our model for $n=5$, 
lies in the range between $8 \times 10^{-8}$ and 
$4 \times 10^{-6}$,
which is consistent with the data in natural population. 

In conclusion, we have presented a model for evolving population.
It has been revealed that mutation in the population
provides a variety in the gene pool and leads to genetic polymorphism.
Such polymorphism, which appears commonly in natural population, 
plays a key role for survival of the population in
the changing environment.
It is thus concluded that 
through mutation, the population can adapt to the changing environment,
maintaining genetic polymorphism.
The predicted proportions of polymorphic and heterozygous loci
as well as the mutation rate
are consistent with the data observed in nature.
The genetic polymorphism has also been discussed from the viewpoint of 
the Hardy-Weinberg theorem.

This work was supported in part by the Basic Sciences Research Institute
Program, Ministry of Education of Korea and in part by the Korea Science
and Engineering Foundation through the SRC Program.

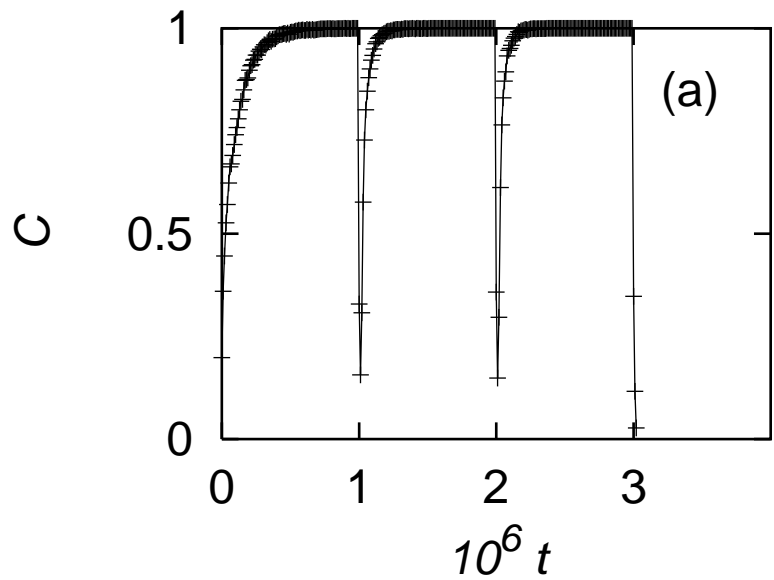
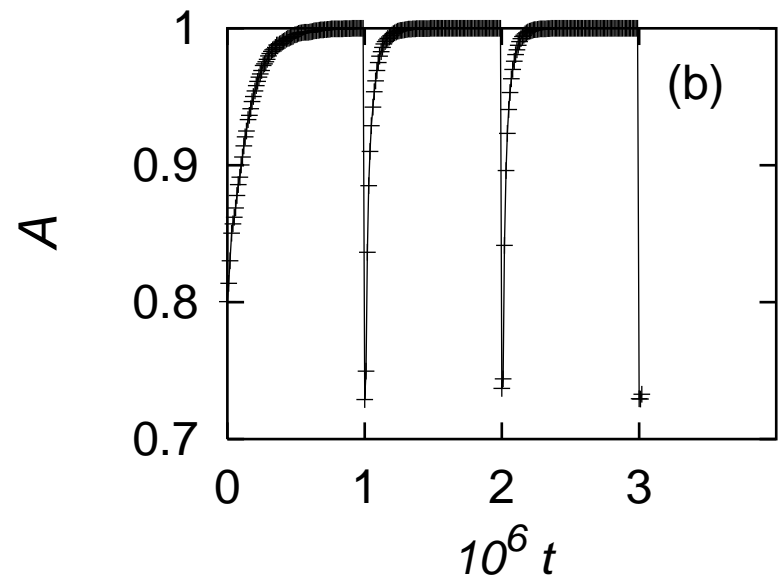
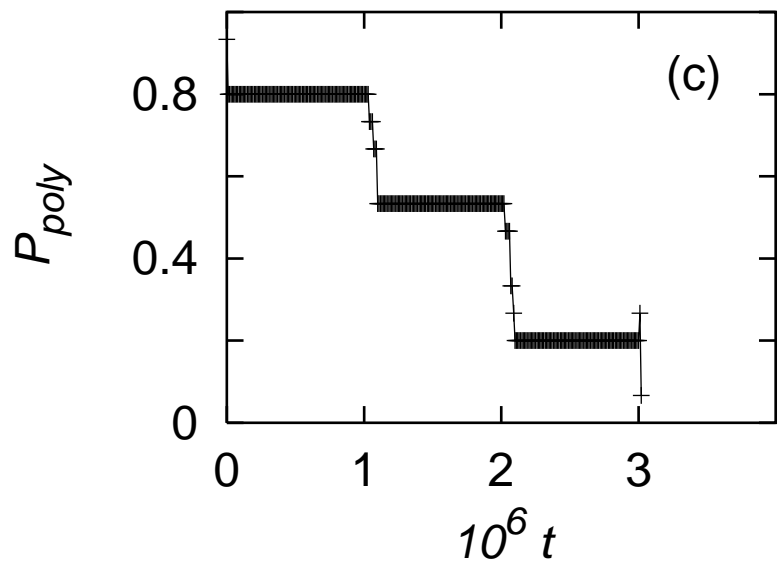
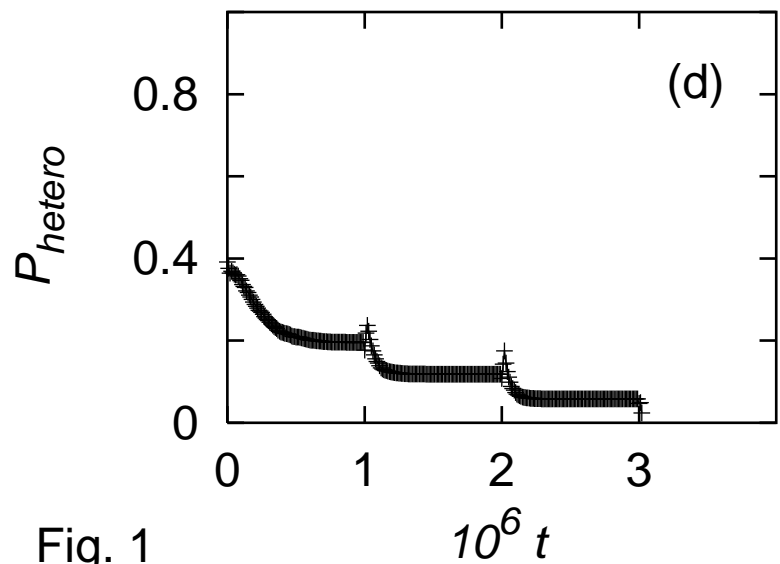
\begin{figure}
\caption{Behavior of
(a) the concentration of surviving population, 
(b) the average adaptation, 
(c) the proportion of polymorphic loci, 
and (d) the proportion of heterozygous loci, 
as functions of time $t$ in units of the generation,
under the environmental change.
In each change, $n{=}4$ loci have been changed among the total $N{=}15$ loci.
The initial values $M_0=1000$ and $C_0=0.2$ have been chosen.
The system is shown to become monomorphic, leading to complete extinction.
Lines are guides to the eye.}
\end{figure}

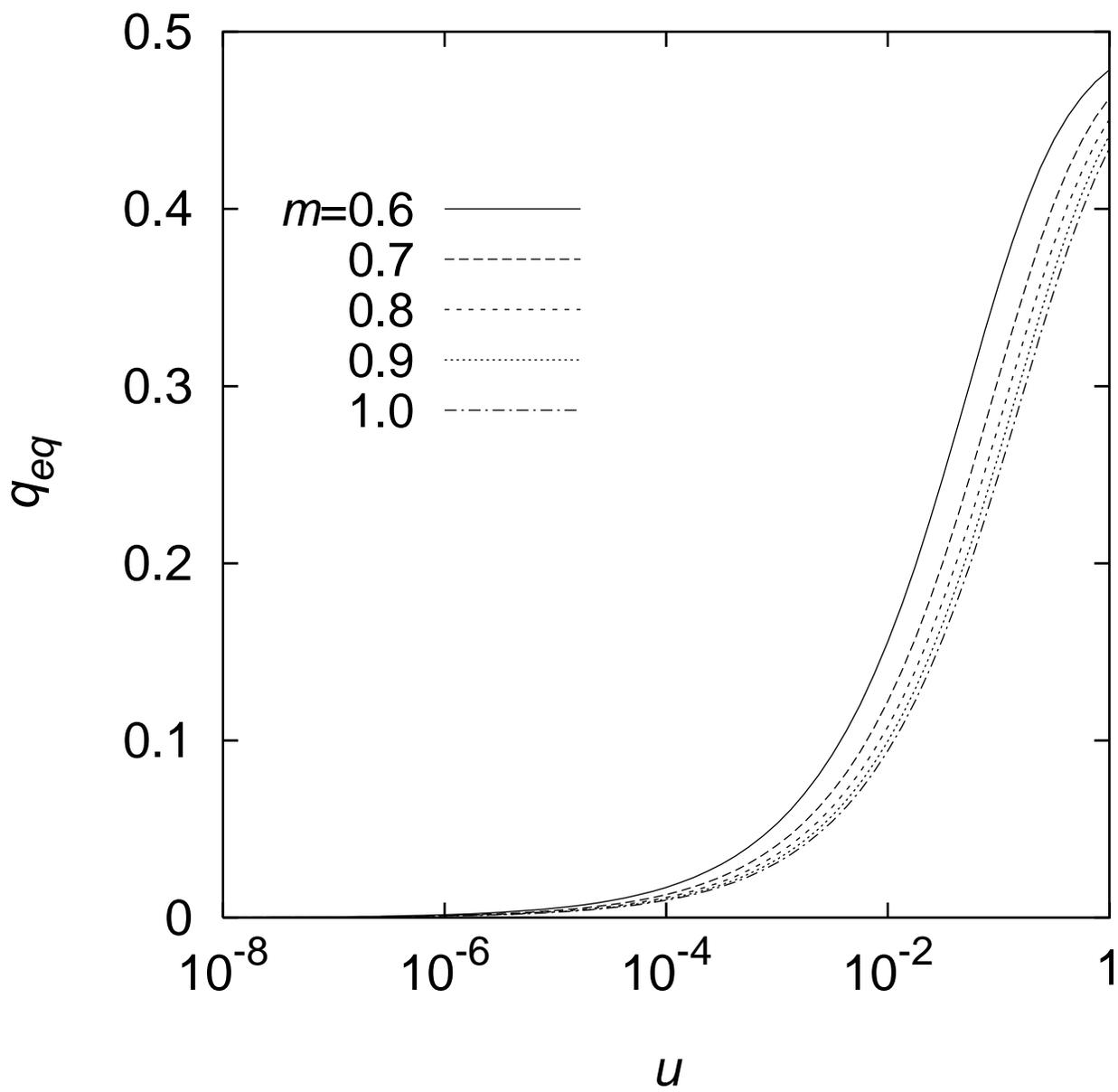
\begin{figure}
\caption{ The equilibrium frequency $q_{eq}$ of the recessive allele
as a function of the mutation rate $u$ per generation, 
for various values of $m$.
Regardless of $m$, the equilibrium frequency $q_{eq}$ increases 
with $u$, preventing monomorphic behavior.}
\end{figure}

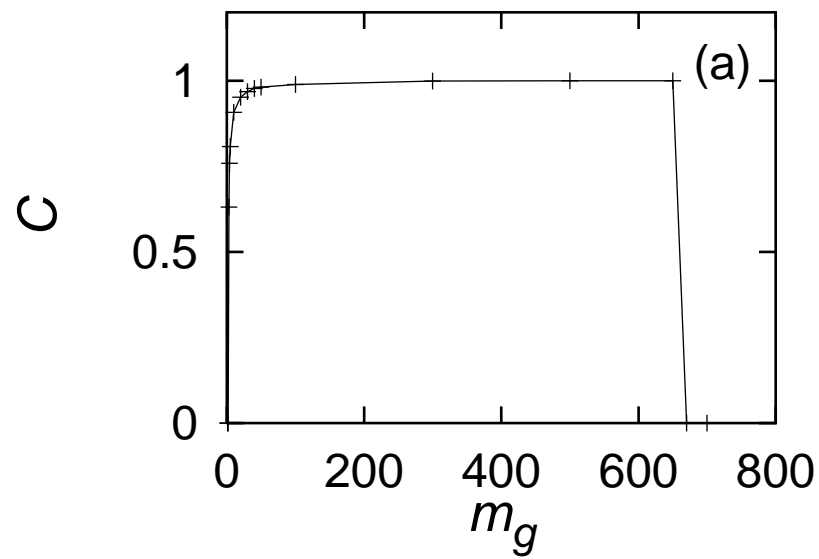
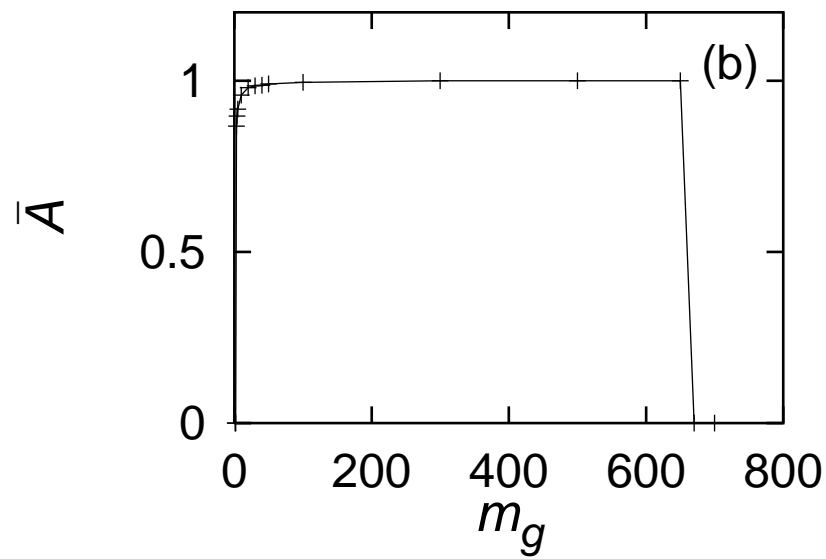
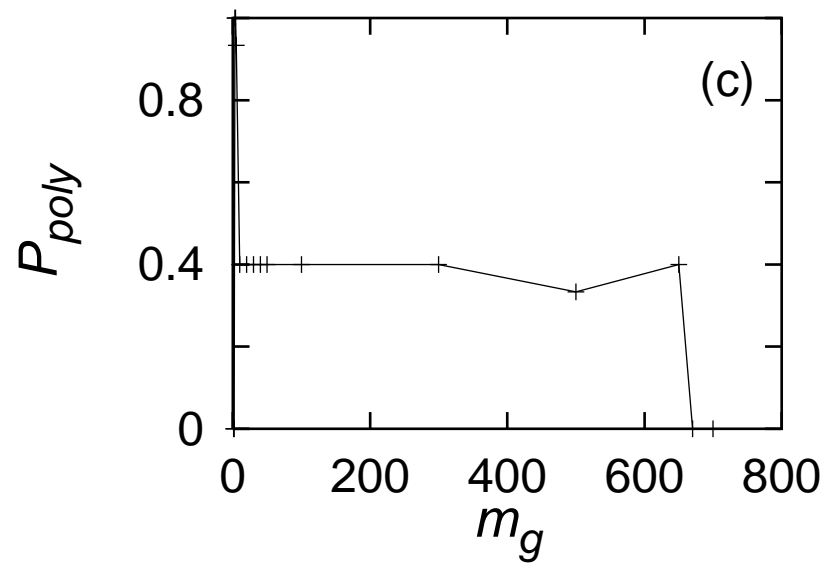
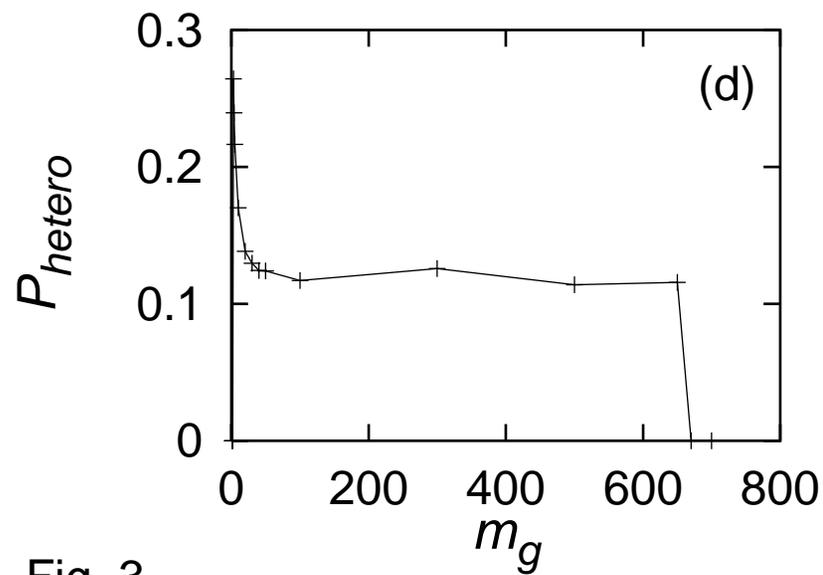
\begin{figure}
\caption{Effects of mutation in the same system as in Fig.\ 1:
(a) the concentration of surviving population, (b) the average adaptation, 
(c) the proportion of polymorphic loci, 
and (d) the proportion of heterozygous loci
as functions of the mutation interval $m_g$ in units of the generation,
measured at time $t=10^7$ generations after ten environmental changes.
In the appropriate range of the mutation rate, 
the population survives the environmental change.
Lines are guides to the eye.}
\end{figure}

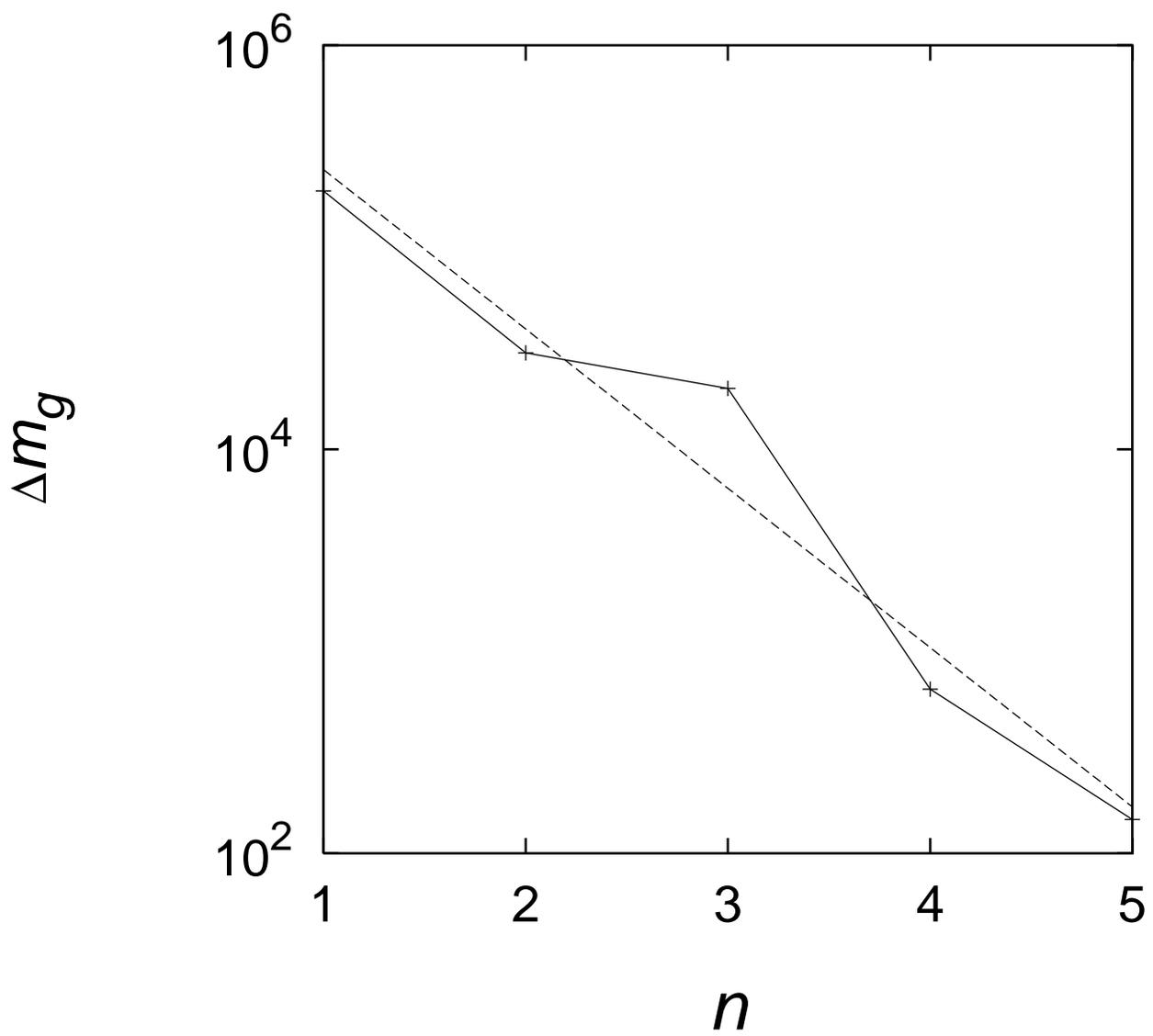
\begin{figure}
\caption{ The range of the mutation interval for survival of population,
$\Delta m_g$, as a function of $n$.
While the solid line is a guide to the eye, 
the dashed line represents the least-square fit with the 
slope $-1.81\pm$ 0.18.  It is shown that
$\Delta m_g$ tends to shrink exponentially with $n$ and vanishes
at $n=6$.}
\end{figure}

\end{multicols}
\end{document}